# Imaging strain-localized exciton states in nanoscale bubbles in monolayer WSe$_2$ at room temperature

Thomas P. Darlington[1], Christian Carmesin[2], Matthias Florian[2], Emanuil Yanev[3], Obafunso Ajayi[3], Jenny Ardelean[3], Daniel A. Rhodes[3], Augusto Ghiotto[4], Andrey Krayev[5], K. Watanabe[6], T. Taniguchi[6], Jeffrey W. Kysar[3], Abhay N. Pasupathy[4], James C. Hone[3], Frank Jahnke[2], Nicholas J. Borys[7,*], and P. James Schuck[3,†]

[1]Department of Physics, UC Berkeley, Berkeley, CA, United States.
[2]Institute for Theoretical Physics, University of Bremen, Bremen, Germany.
[3]Department of Mechanical Engineering, Columbia University, New York, NY, United States.
[4]Department of Physics, Columbia University, New York, NY, United States.
[5]Horiba Scientific, Novato, CA, United States.
[6] National Institute for Materials Science, Tsukuba, 305-0047, Japan.
[7]Department of Physics, Montana State University, Bozeman, MT, United States.

**Abstract:**
In monolayer transition metal dichalcogenides, quantum emitters are associated with localized strain that can be deterministically applied to create designer nano-arrays of single photon sources. Despite an overwhelming empirical correlation with local strain, the nanoscale interplay between strain, excitons, defects and local crystalline structure that gives rise to these quantum emitters is poorly understood. Here, we combine room-temperature nano-optical imaging and spectroscopy of excitons in nanobubbles of localized strain in monolayer WSe$_2$ with atomistic structural models to elucidate how strain induces nanoscale confinement potentials that give rise to highly localized exciton states in 2D semiconductors. Nano-optical imaging of nanobubbles in low-defect monolayers reveal localized excitons on length scales of ~10 nm at multiple sites along the periphery of individual nanobubbles, which is in stark contrast to predictions of continuum models of strain. These results agree with theoretical confinement potentials that are atomistically derived from measured topographies of existing nanobubbles. Our results provide one-of-a-kind experimental and theoretical insight of how strain-induced confinement—without crystalline defects—can efficiently localize excitons on length scales commensurate with exciton size, providing key nanoscale structure-property information for quantum emitter phenomena in monolayer WSe$_2$.

The intense light-matter interactions of two-dimensional (2D) monolayer transition metal dichalcogenides (1L-TMDs) are mediated by a diverse suite of excitonic phenomena that present a wealth of opportunities for novel optoelectronic functionalities in areas spanning from high-performance sensing and non-traditional photovoltaics to the quantum information sciences. Many of these opportunities emerge from—and heavily rely on—the unique ways in which the 2D TMD semiconductors enable the manipulation of excitonic phenomena on the nanoscale. Within this quiver of capabilities[1-8], strain engineering is preeminent, offering unprecedented flexibility and precision, and opening new routes for highly tailored optoelectronic materials. Extrinsic strains of up to several percent can be endured without fracture[9-11] and have been shown to continuously reduce the optical bandgap[9,12] as well as modify the exciton-phonon coupling, thus narrowing the photoluminescence (PL) linewidth[13]. Further, such strain can be localized, embedding nanoscale potential energy wells for the funneling and localization of excitons[14-16]. Notably, in 1L-WSe$_2$, it has become evident that local strain is a key ingredient for the formation of low-temperature quantum emitters[11,17,18], and that strain-engineering can potentially extend their operating range to room-temperature[19,20]. The resulting highly integrable solid-state non-classical light sources[21] underscore the remarkable technological potential of harnessing nanoscale strain engineering of exciton localization in 2D TMD semiconductors.

---

[*] nicholoas.borys@montana.edu
[†] p.j.schuck@columbia.edu

Despite the observations of exciton funneling and single-photon emission in localized strained regions of 2D TMDs, an understanding of the induced exciton localization is notably lacking, especially on the nanoscale. This absence of a fundamental picture has led to critical ambiguities, particularly for the formation of quantum emitters in 1L-WSe$_2$ where the roles and interplay between strain, excitons, and crystallographic defects remain largely a mystery. For instance, it has been shown that the quantum emission can be co-localized with a nanobubble[11] where strain models derived from continuum elastic plate theory predict that a single region of maximum strain (and thus exciton localization) occurs at the apex of the nanobubble[9,10]. However, multiple emitters per nanobubble are typically observed, which is at odds with a single localization site and suggests a possible role of crystallographic defect states or some other sub-nanobubble inhomogeneity[11,18,22-25]. Recently, improved microscopic theoretical models predict that quantum-dot-like electronic states in TMDs can form within nanobubbles in the absence of defects[26,27]. In particular, a first-principles approach that carefully considers the strain-induced atomic structure[27] (and which we employ here) predicts that strain maxima and multiple low-energy states form in a doughnut-like distribution near the nanobubble periphery. This surprising strain distribution is an apparent result of atomic scale wrinkling around the edges of highly-strained nanobubbles, leading to more localized lower energy states relative to the case of a nanobubble with a smooth topography. Such predicted states provide a critical missing puzzle piece for understanding and controlling strain-localization of excitons in 2D TMD semiconductors. However, due to their nanoscale size and distribution within a nanobubble, these localized energy states cannot be directly resolved with far-field optical characterization, necessitating interrogation by more sophisticated higher-resolution techniques.

In this work, we employ apertureless scanning near-field optical microscopy to image the localized exciton (LX) states within nanobubbles in 1L-$WSe_2$. Using hyperspectral nano-photoluminescence (nano-PL) mapping, we achieve a sub-20 nm spatial resolution and resolve individual localized low-energy states that are separated by distances less than 50 nm within single nanobubbles at room temperature. Furthermore, we observe that these localized states form in doughnut-like patterns that are consistent with the first-principles predictions[27] that refine those of continuum plate and membrane models[9,10]. These LXs are also identified in nanobubbles of "flux-grown" 1L-$WSe_2$, which has a much lower defect density as compared to monolayers exfoliated from commercially available crystals (ca. 100× lower defect density)[28]. Our results are consistent and reproducible across multiple samples and numerous bubbles (N>50; see, for example,

Supplementary Fig. 3). The combination of nano-optical characterization with atomistic modeling and materials engineering indicates that (i) exciton localization by inhomogeneous strain occurs via the formation of wrinkles near the edges and bending near the base of nanobubbles; (ii) that these excitons remain localized at elevated (room) temperature; and (iii) the optical emission of LXs at room temperature can be enhanced by nano-optical/plasmonic techniques by at least 100×. Together, these findings provide key experimental evidence of highly-confined localized states in nanobubbles, which constitute a new paradigm of nanoscale strain engineering in 2D TMD semiconductors that is particularly relevant for developing non-classical light sources for practical quantum optical devices.

Numerous processes that are critical for developing advanced functionality and novel devices with 1L-TMD semiconductors have been studied with nano-optical methods, revealing a rich suite of highly localized optoelectronic phenomena[29-34]. Figure 1a illustrates the experimental configuration of our nano-PL investigations of individual nanobubbles of 1L-WSe$_2$ where a sharp silver tip (Horiba Scientific) has been positioned within 2 nm of the 1L-WSe$_2$ (Fig. 1a). The tip-sample junction was illuminated from the side with *p*-polarized, continuous wave laser excitation (637.27 nm; 75-100 µW) at an oblique angle of incidence such that the induced polarization of the tip can couple to the in-plane and out-of-plane transition dipoles[35]. As the sample is raster scanned, nano-PL from the tip-sample junction and local topography of the 1L-WSe$_2$ are simultaneously recorded.

As indicated in the atomic force microscopy (AFM) micrograph (Fig. 1b), the entirety of the 1L-WSe$_2$ in Figs 1-3 is supported by an underlying crystal of hexagonal boron nitride (hBN; ~200 nm thick) forming a 2D 1L-WSe$_2$/hBN heterostructure. We note that a portion of this 1L-WSe$_2$ flake is fully encapsulated and covered by a second, thin hBN crystal (~3 nm). However, all of the results presented here were acquired from the exposed 1L-WSe$_2$ that is directly accessible to the nano-optical antenna (as illustrated in Fig. 1a). The details of the "dry-stamping" sample fabrication are provided in earlier work[11] and Supplementary Information. This process is known to produce structural "imperfections" such as nanobubbles, tears, folds and wrinkles. In our case, a portion of the 1L-WSe$_2$ is folded on top of itself to form a bilayer and several pronounced creases (Fig. 1b). In addition, small topographical nanobubbles, with characteristic heights of ~2-10 nm, are observed in the 1L-WSe$_2$ and are attributed to trapped substances between the layers[10]. Such nanobubbles have been repeatedly observed in heterostructures of layered materials[11,15,36], and as noted above are known to be correlated with quantum emitters in 1L-WSe$_2$ at cryogenic temperatures[11].

Figure 1c contrasts nano-PL spectra collected from a flat region and a nanobubble of the 1L-WSe$_2$ acquired at room temperature. On the flat region, the spectrum is composed of the standard excitonic PL of 1L-WSe$_2$ at ~1.65 eV,[29,37] which we will refer to as the primary exciton (PX). When the nano-optical antenna is positioned over a nanobubble, the emission spectrum shows a dramatic change, exhibiting red-shifted excitonic emission and an additional intense low-energy emission band that is centered at ~1.56 eV, as determined by Gaussian peak fitting (Supplementary Fig. 2 and Supplementary Methods). In the 1L-WSe$_2$, every region that was found to exhibit strong low-energy emission was correlated with a nanobubble. A selection of such emission spectra from four individual nanobubbles and their associated topographies is shown in Supplementary Figure 1. The lateral extents of the nanobubbles range from ~30–140 nm with heights from 1.5-20 nm, and depending on size, exhibit a diverse set of broad emission spectra, some of which show signatures of multiple low-energy states. In general, low-energy emission occurs at emission energies that span a range of about 70 meV, from ~80 meV to ~150 meV below the PX, and exhibits linewidths of ~150 meV that are attributed to thermal broadening[20] and plasmonic coupling to the tip[38].

Close inspection of the nano-PL spectra in Fig. 1c reveals several characteristics of the low-energy PL that are different from the well-studied PX state. First, we concurrently observe both the low-energy emission and strong emission at ~1.65 eV, corresponding to the PX spectra measured in the far-field (inset Fig. 1c), within our nano-optical mode volume. Second, the low-energy band here is well below the energies of other native excitonic complexes in this material (e.g. trions, the dark exciton[39,40], and biexcitons[41]) and in an energetic region often associated with defect states[41]. Third, the linewidth is substantially larger than that of the PX, which contrasts with what has been observed for strain-tuned PX states[13]. And finally, the energy separation of this band with respect to the PX state (see also Supplementary Fig. 2) and its localization to nanobubble regions are in-line with corresponding observations for single emitter states at cryogenic temperatures in 1L-WSe$_2$[11,18,23-25,42,43]. Based on these considerations, we deduce that this low-energy band, which is amplified significantly by the nano-optical antenna (consistent with the observation in ref. [20] of plasmon-enhanced localized emission at elevated temperatures), originates from LX states in the nanobubbles.

The nano-PL directly probes the LX emission on length scales that are commensurate with – and even smaller than – the nanobubbles themselves, allowing us to explore the origins of the diverse emission spectra and nanoscale structure-property relationships in more detail. Here, we estimate that our spatial resolution is ~15 nm. In Figure 2, the spatial distribution of LX emission is shown for a larger nanobubble (approximate radius of 75 nm; peak height of 12 nm; aspect ratio of 0.16). With nano-PL, we resolve nanoscale variations of the LX emission within the nanobubble itself, finding that the integrated emission intensity is not uniform (integrated over the LX spectral region of 1.5-1.6 eV; Fig. 2a): it is concentrated to specific locations of the nanobubble that do not clearly correspond to its apex (cf. inset of Fig. 2a). Furthermore, different points of the nanobubble separated by distances as small as 30 nm exhibit clearly distinct spectra (Fig. 2b). The dashed curve is drawn as guide to the eye between the maxima of LX spectral peaks, clearly showing that the redder LX states are positioned to the edges of the nanobubble, where the maximum confinement potential and possible wrinkle formation are expected to occur. Clearly, larger nanobubbles host multiple emissive localization centers for LXs, an effect that to date has only been inferred spectroscopically[11,42].

Our hyperspectral mapping of the nanobubble in Fig 2 shows that the lowest-energy states (and thus deepest confinement potentials) are concentrated around the edge. To gain more insight into the physical mechanism, we applied the theory of ref. [27], which considers the influence of local strain within the nanobubbles on electronic and optical properties. Our method is based on atomistic calculations, in which the relevant part of the nanostructure is modeled using a lattice of atomic sites. For the present calculations, we use the nanobubble geometry obtained from AFM measurements and determine the corresponding relaxed positions of the individual atoms in a valence force field simulation. The result provides the strain field in the structure and confirms the existence of atomic-scale wrinkling. In a second step, the information about the displaced atomic positions is used within a tight-binding calculation in order to quantify how the strain field of the individual atoms within the nanobubble translates into a confinement potential and local electronic states.

For the nanobubble topography given by the AFM data of Fig. 2a (inset), the calculated confinement potential (Fig. 3a) exhibits a doughnut pattern, with deeper potentials located on the nanobubble periphery that traces the wrinkling effect (Fig. 3b) and corresponds remarkably well to the nanoscale spatial distribution of measured LX energies (Fig. 3c; see Supplementary Information for fitting details). This correspondence is further in direct accordance with the experimental spectra in Figure 2b, which shows that the low-energy emission is isolated to the nanobubble periphery. To demonstrate the correlation between the theoretical predictions and experimental data, Fig. 3d shows the measured LX energy plotted against the predicted confinement potential at the corresponding region of the nanobubble (see Supplementary Information for analysis details). Around the periphery of the nanobubble, a strong correlation between the LX energy and the depth of the confinement potential is observed: regions of stronger confinement correspond to lower LX energies. In contrast, in the center region, very little correlation of energy with confinement potential is observed, which suggests a lack of strongly confined excitons in that area (masks for each of these regions is shown in Supplementary Fig. 4). This different behavior may indicate a more-general exciton funneling effect in the central region[14], whereas the periphery is dominated by highly confined states.

The correspondence between the theoretical analysis of the strain and resulting electronic states within the nanobubble and our nano-optical dataset presented in Fig. 3 strongly suggests that the experimentally localized states observed are the predicted quantum-dot-like states[27]. In principle, point defects in the 1L-WSe$_2$ lattice can also localize excitons in a similarly inhomogeneous way. To better elucidate the potential role of defects, we repeated our nanobubble PL mapping and theoretical calculations in high-quality flux-grown 1L-WSe$_2$, which has been shown to have defect densities approximately two orders of magnitude lower than commercially grown crystals–providing for the first time an average spacing between defects on the order of nano-optical resolution[28]. As shown in detail in the Supplementary Information, every nanobubble identified in this higher-quality material exhibits similar LX emission to that of the commercial material (see Supplementary Fig. 3 and corresponding discussion). More specifically, the occurrence of the LX emission band in the nanobubbles does not depend on the defect density in the ranges probed here.

In Figure 4, we provide a collage of four nanobubbles in the monolayer area of the high-quality (i.e., flux-grown) 1L-WSe$_2$. Row I of Figure 4 contains the measured AFM topography for the four nanobubbles with their corresponding spatial distributions of LX emission energies using 785 nm excitation shown in Row II. Exciting with 785 nm light, which is below the energy of the PX, enhances the contrast between the LX and PX states, allowing the LX states to be observed with a greater signal-to-noise ratio (see Supplementary Information). Rows III, IV, and V summarize the theoretical results of the confinement potentials, overlap of the electron-hole wavefunctions, and the predicted shifts in PL emission energies, respectively. As with the nanobubble shown in Fig. 3b, each nanobubble in the flux-grown 1L-WSe$_2$ exhibits lower energy emission concentrated at the nanobubble periphery. This peripheral spatial distribution of low-energy states is replicated in the predicted confinement potentials and electron-hole wavefunction overlap shown Rows III and IV of Fig. 4, respectively. Finally, the predicted PL energy shifts and corresponding spatial distributions shown in Row V agree well with the measured LX emission energy (Row II). Considering the challenges of room-temperature hyperspectral imaging at the 10-20 nm length scale, limitations of the topographical maps in Row I to a resolution of ~15 nm, as well as the complexities to directly incorporate these measurements into a feasible atomistic model (for details, see Supplementary Information), the agreement between experiment and theory is particularly remarkable.

The similarities between Figures 3 and 4—despite dramatically different defect densities—suggest that LX emission distribution in nanobubbles is not significantly determined by atomic defects in the lattice. As noted above, since there is approximately one defect every 15 nm × 15 nm on average in flux-grown material – near the spatial resolution of the nano-optical probe – then major defect-associated optical changes would be resolvable. While defects may still play a role in the emission, the implication here is that the primary effect in terms of the nanoscale spatial distribution and creation of quantum-dot-like states is the local strain profile, with localized maximum strain occurring at locations around the edge of nanobubbles leading to LX emission observable at room temperature with a nano-optical probe. Still, it is certainly possible that defects play a role in e.g. further localization or the exact distribution of wrinkles at the atomic scale. Such angstrom-level effects are beyond the resolution of our current nano-optical probes, which utilize AFM tips with radii of curvature ranging from ~15 to ~40 nm. Future correlations with e.g. atomic-resolution AFM[44] should be able to map these effects.

In conclusion, using nano-optical techniques, strain localized exciton states within single nanobubbles of 1L-WSe$_2$ are observed at room temperature with localization lengths scales of less than 50 nm and significantly smaller than the nanobubble itself. In general, the spatial distribution of these low-energy states forms a doughnut-like pattern around the periphery of the nanobubble that surrounds a region of higher-energy states at its center, which is consistent with previous theoretical predictions of atomic-scale strain effects. By integrating the experimentally determined topography with this atomistic model, the experimental results are qualitatively replicated by the theoretical model, establishing the first robust experiment-theory connection of the formation of strain-induced quantum-dot states in monolayer TMD semiconductors. Furthermore, we reproduce these findings in 1L-WSe$_2$ with orders-of-magnitude lower defect densities, implying the nanoscale localization is primarily influenced by strain and not crystalline defects. Our results show that LX states accessed with nano-optical techniques remain localized at room temperature, leading to compelling potential applications in practical strain-engineered optoelectronic and quantum-optical architectures.

**Acknowledgements:** NJB and PJS gratefully acknowledge support from the National Science Foundation under award NSF-1838403. PJS thanks Prof. S. Strauf for stimulating and enlightening discussions. JWK gratefully acknowledges support from the National Science Foundation under awards NSF-1437450 and NSF-1363093. Preparation and characterization of nanobubbles in the flux-grown low-defect-density $WSe_2$ is partially supported as part of Programmable Quantum Materials, an Energy Frontier Research Center funded by the U.S. Department of Energy (DOE), Office of Science, Basic Energy Sciences (BES), under award DE-SC0019443. CC, MF, and FJ gratefully acknowledge computational resources from HLRN Berlin and funding from the German Science Foundation (DFG) via the graduate school "Quantum-Mechanical Material Modeling". KW and TT acknowledge support from the Elemental Strategy Initiative conducted by the MEXT, Japan and the CREST (JPMJCR15F3), JST.

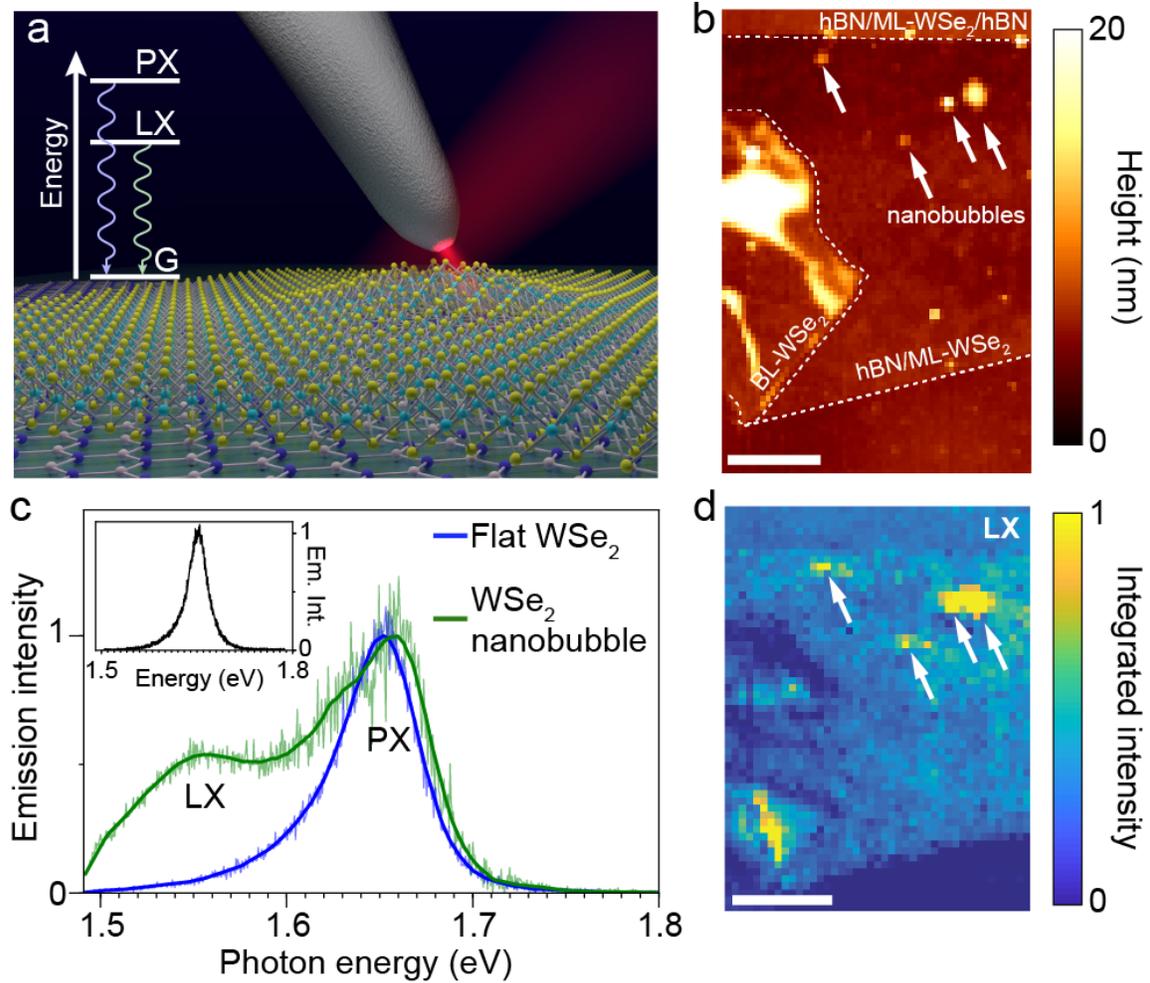

**Figure 1** – nano-optical detection of room-temperature photoluminescence from localized\ exciton (LX) states localized to nanobubbles in 1L-WSe2. (a) Schematic of the room-temperature nano-PL imaging and spectroscopy of 1L-WSe2 on top of a hBN. Inset: simple level diagram illustrating the energetic ordering between the primary exciton (PX), LX and ground (G) states of 1L-WSe2. (b) AFM topography of the 1L-WSe2 flake exfoliated on top of the hBN substrate. The white arrows mark several nanobubbles in the 1L-WSe2. (c) Comparison of nano-PL emission spectra collected from the 1L-WSe2 on and off a nanobubble with a resolution of 20 nm. Inset: typical far field spectrum of the 1L-WSe2. (d) Spatial map of the LX emission (integrated from 1.5-1.6 eV) showing that the LX emission in the 1L-WSe2 is localized to spatially discrete regions that correspond to nanobubbles in the topography. Inset: similar map of the PX state showing diminished intensity in the bilayer region. All scale bars are 500 nm.

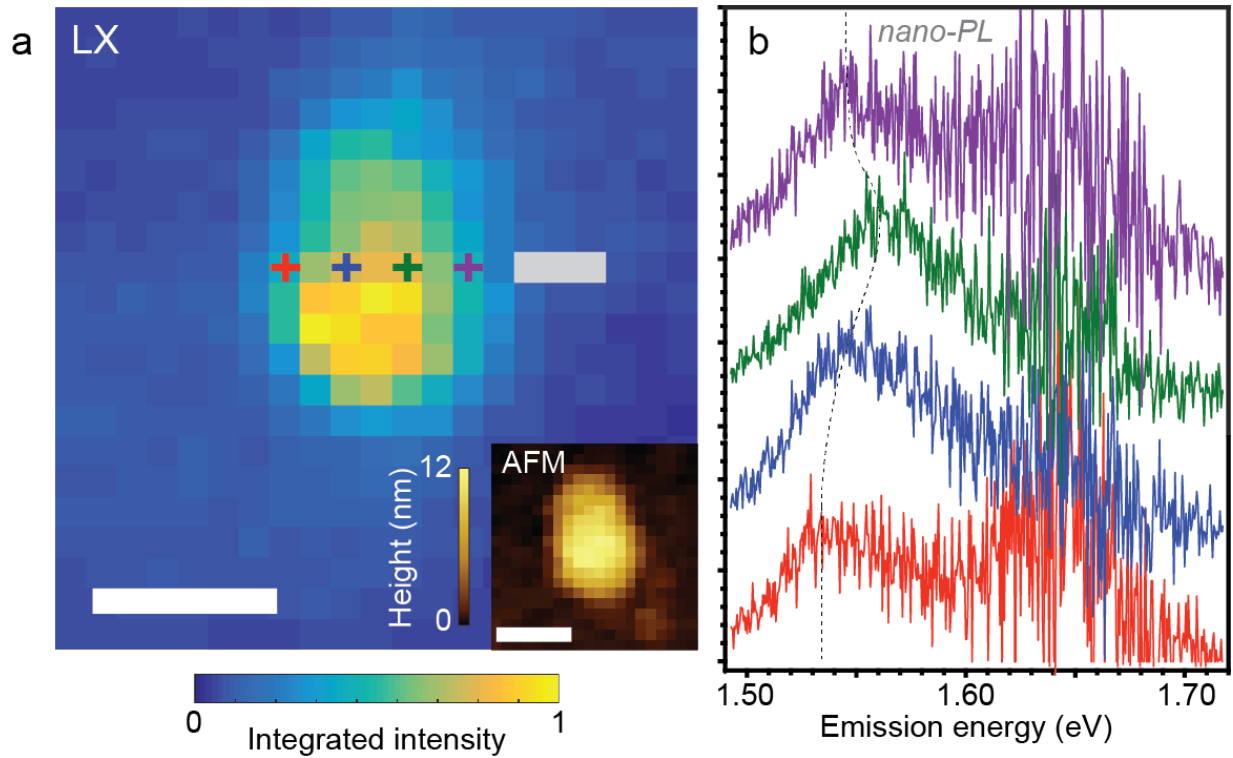

**Figure 2** - high-resolution nano-PL imaging and spectroscopy of distinct LX localization centers within a single nanobubble. (a) Nano-PL image of the spatial distribution of LX emission (integrated intensity from 1.5-1.6 eV) of a single nanobubble. The three grey pixels are a scan artifact where nano-optical signal was not acquired. Scale bar: 100 nm. Inset is the topography of the nanobubble (scale bar: 100 nm). Its overall height is 12 nm and its average radius is 74 nm. (b) Sample emission spectra from the corresponding points labeled in panel (a). The distance between the red and blue marks is 45 nm.

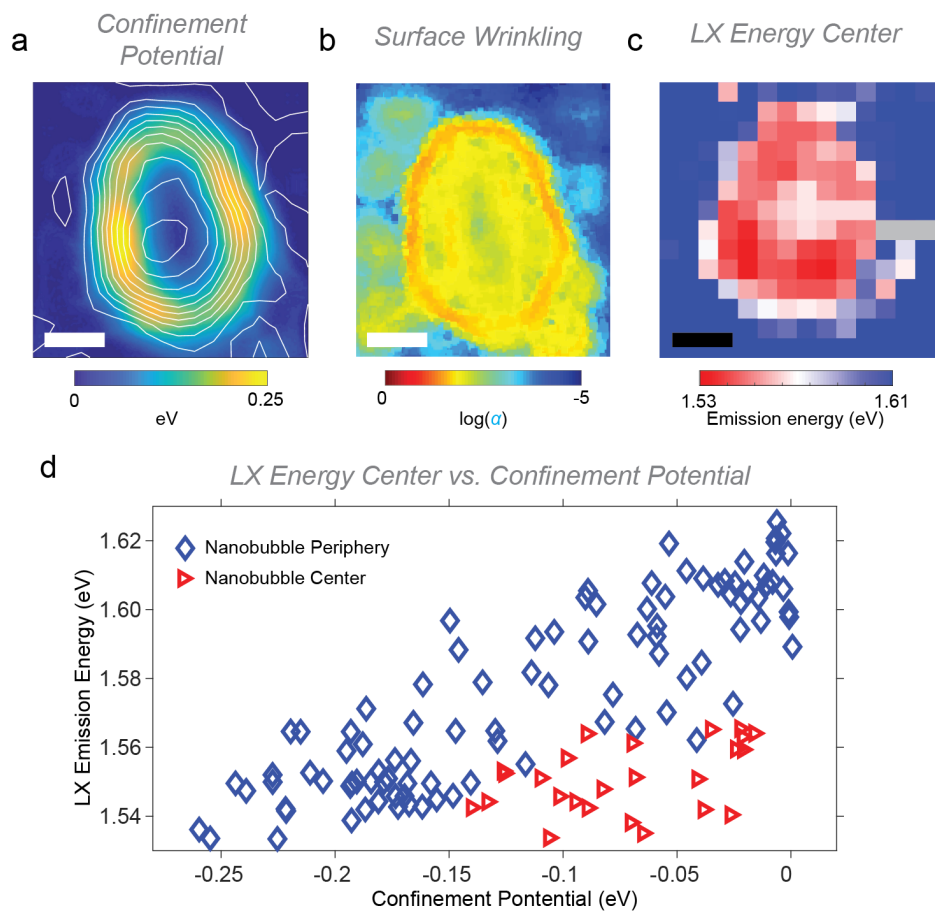

**Figure 3** – (a) Calculated carrier confinement potential for the nanobubble topography shown in Fig. 2a inset. The solid lines represent topographic contours based on the AFM measurement. (b) Corresponding values of the surface normal deviation, defined as the average angle $\alpha$ between normal vectors at neighboring unit cells, obtained from relaxed atomic positions, showing wrinkling of the unit cell. (c) Map of experimentally measured LX emission energy from the hyperspectral nano-PL shown in Fig 2. (d) Correlation between the local confinement potentials in (a) and LX emission energy shown in (c). The datapoints are separated into points from the center of the nanobubble (red triangles) and points from the periphery of the nanobubble (blue diamonds). The scale bar of the experiment in (c) is 50 nm, while theory calculations with atomic resolution are scaled up by 5x (see Supplementary Information).

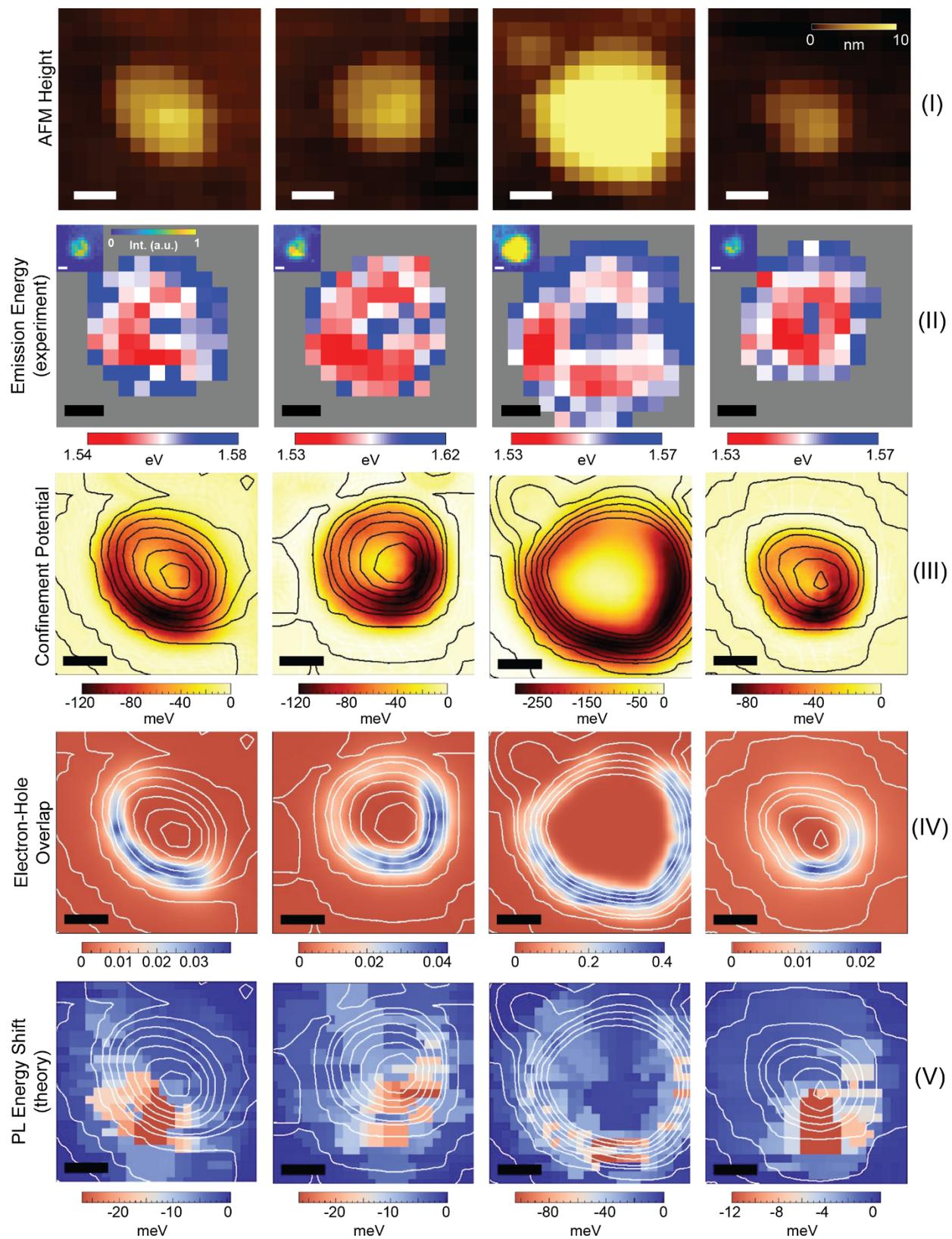

**Figure 4** – Comparison of experiment and theory for 4 nanobubbles in low-defect monolayer WSe$_2$. Rows I and II show the measured AFM topography and nano-PL emission energy, respectively. The inset of row II shows the integrated PL intensity after a low pass energy filter to isolate the LX emission. Row III provides theoretical results of the confinement potential obtained from displaced atomic positions using nanobubble strain calculations and the Harrison rule[47]. Rows IV and V contain electron and hole wavefunction overlap, and predicted PL energy shift, respectively. The solid lines in rows III-V represent topographic contours using the nanobubble topographies shown in Row 1. Scale bars are 50 nm in Row I and II. Theory calculations with atomic resolution are scaled up by 5x (see Supplementary Information).

**References:**

1. Yao, K. *et al.* Optically Discriminating Carrier-Induced Quasiparticle Band Gap and Exciton Energy Renormalization in Monolayer MoS$_2$. *Physical Review Letters* **119**, 087401, doi:10.1103/PhysRevLett.119.087401 (2017).
2. Mak, K. F. *et al.* Tightly bound trions in monolayer MoS$_2$. *Nature Materials* **12**, 207, doi:10.1038/nmat3505 (2013).
3. Raja, A. *et al.* Coulomb engineering of the bandgap and excitons in two-dimensional materials. *Nature Communications* **8**, 15251, doi:10.1038/ncomms15251 (2017).
4. Unuchek, D. *et al.* Room-temperature electrical control of exciton flux in a van der Waals heterostructure. *Nature* **560**, 340, doi:10.1038/s41586-018-0357-y (2018).
5. Alexeev, E. M. *et al.* Resonantly hybridized excitons in moire superlattices in van der Waals heterostructures. *Nature* **567**, 81, doi:10.1038/s41586-019-0986-9 (2019).
6. Jin, C. H. *et al.* Observation of moire excitons in WSe$_2$/WS$_2$ heterostructure superlattices. *Nature* **567**, 76, doi:10.1038/s41586-019-0976-y (2019).
7. Seyler, K. L. *et al.* Signatures of moire-trapped valley excitons in MoSe$_2$/WSe$_2$ heterobilayers. *Nature* **567**, 66, doi:10.1038/s41586-019-0957-1 (2019).
8. Tran, K. *et al.* Evidence for moire excitons in van der Waals heterostructures. *Nature* **567**, 71, doi:10.1038/s41586-019-0975-z (2019).
9. Lloyd, D. *et al.* Band Gap Engineering with Ultralarge Biaxial Strains in Suspended Monolayer MoS$_2$. *Nano Letters* **16**, 5836-5841, doi:10.1021/acs.nanolett.6b02615 (2016).
10. Khestanova, E., Guinea, F., Fumagalli, L., Geim, A. K. & Grigorieva, I. V. Universal shape and pressure inside bubbles appearing in van der Waals heterostructures. *Nature Communications* **7**, 12587, doi:10.1038/ncomms12587 (2016).
11. Shepard, G. D. *et al.* Nanobubble induced formation of quantum emitters in monolayer semiconductors. *2D Materials* **4**, doi:10.1088/2053-1583/aa629d (2017).
12. Desai, S. B. *et al.* Strain-induced indirect to direct bandgap transition in multilayer WSe$_2$. *Nano Letters* **14**, 4592-4597, doi:10.1021/nl501638a (2014).
13. Niehues, I. *et al.* Strain Control of Exciton–Phonon Coupling in Atomically Thin Semiconductors. *Nano Letters* **18**, 1751-1757, doi:10.1021/acs.nanolett.7b04868 (2018).
14. Feng, J., Qian, X., Huang, C.-W. & Li, J. Strain-engineered artificial atom as a broad-spectrum solar energy funnel. *Nature Photonics* **6**, 866-872, doi:10.1038/nphoton.2012.285 (2012).
15. Tyurnina, A. V. *et al.* Strained Bubbles in van der Waals Heterostructures as Local Emitters of Photoluminescence with Adjustable Wavelength. *ACS Photonics* **6**, 516-524, doi:10.1021/acsphotonics.8b01497 (2019).

16	Li, H. *et al.* Optoelectronic crystal of artificial atoms in strain-textured molybdenum disulphide. *Nature Communications* **6**, 7381, doi:10.1038/ncomms8381 (2015).
17	Branny, A., Kumar, S., Proux, R. & Gerardot, B. D. Deterministic strain-induced arrays of quantum emitters in a two-dimensional semiconductor. *Nature Communications* **8**, 15053, doi:10.1038/ncomms15053 (2017).
18	Tonndorf, P. *et al.* Single-photon emission from localized excitons in an atomically thin semiconductor. *Optica* **2**, doi:10.1364/optica.2.000347 (2015).
19	Rosenberger, M. R. *et al.* Quantum Calligraphy: Writing Single-Photon Emitters in a Two-Dimensional Materials Platform. *ACS Nano* **13**, 904-912, doi:10.1021/acsnano.8b08730 (2019).
20	Luo, Y., Liu, N., Li, X., Hone, J. C. & Strauf, S. Single photon emission in $WSe_2$ up 160 K by quantum yield control. *2D Materials* **6**, doi:10.1088/2053-1583/ab15fe (2019).
21	Aharonovich, I., Englund, D. & Toth, M. Solid-state single-photon emitters. *Nature Photonics* **10**, 631-641, doi:10.1038/nphoton.2016.186 (2016).
22	He, Y. M. *et al.* Single quantum emitters in monolayer semiconductors. *Nature Nanotechnology* **10**, 497-502, doi:10.1038/nnano.2015.75 (2015).
23	Koperski, M. *et al.* Single photon emitters in exfoliated $WSe_2$ structures. *Nature Nanotechnology* **10**, 503-506, doi:10.1038/nnano.2015.67 (2015).
24	Srivastava, A. *et al.* Optically active quantum dots in monolayer $WSe_2$. *Nature Nanotechnology* **10**, 491-496, doi:10.1038/nnano.2015.60 (2015).
25	Chakraborty, C., Kinnischtzke, L., Goodfellow, K. M., Beams, R. & Vamivakas, A. N. Voltage-controlled quantum light from an atomically thin semiconductor. *Nature Nanotechnology* **10**, 507-511, doi:10.1038/nnano.2015.79 (2015).
26	Chirolli, L., Prada, E., Guinea, F., Roldán, R. & San-Jose, P. Strain-induced bound states in transition-metal dichalcogenide bubbles. *2D Materials* **6**, 025010, doi:10.1088/2053-1583/ab0113 (2019).
27	Carmesin, C. *et al.* Quantum-Dot-Like States in Molybdenum Disulfide Nanostructures Due to the Interplay of Local Surface Wrinkling, Strain, and Dielectric Confinement. *Nano Lett.* **19**, 3182-3186, doi:10.1021/acs.nanolett.9b00641 (2019).
28	Edelberg, D. *et al.* Approaching the Intrinsic Limit in Transition Metal Diselenides via Point Defect Control. *Nano Letters* **19**, 4371-4379, doi:10.1021/acs.nanolett.9b00985 (2019).
29	Park, K. D. *et al.* Hybrid Tip-Enhanced Nanospectroscopy and Nanoimaging of Monolayer $WSe_2$ with Local Strain Control. *Nano Letters* **16**, 2621-2627, doi:10.1021/acs.nanolett.6b00238 (2016).
30	Lin, Z. *et al.* 2D materials advances: from large scale synthesis and controlled heterostructures to improved characterization techniques, defects and applications. *2D Materials* **3**, doi:10.1088/2053-1583/3/4/042001 (2016).
31	Kastl, C. *et al.* The important role of water in growth of monolayer transition metal dichalcogenides. *2D Materials* **4**, doi:10.1088/2053-1583/aa5f4d (2017).
32	Bao, W. *et al.* Visualizing nanoscale excitonic relaxation properties of disordered edges and grain boundaries in monolayer molybdenum disulfide. *Nature Communications* **6**, 7993, doi:10.1038/ncomms8993 (2015).
33	Lee, Y. *et al.* Near-field spectral mapping of individual exciton complexes of monolayer $WS_2$ correlated with local defects and charge population. *Nanoscale* **9**, 2272-2278, doi:10.1039/c6nr08813a (2017).
34	Ogletree, D. F. *et al.* Near-Field Imaging: Revealing Optical Properties of Reduced-Dimensionality Materials at Relevant Length Scales. *Advanced Materials* **27**, 5692-5692, doi:10.1002/adma.201570255 (2015).

35  Schuck, P. J., Bao, W. & Borys, N. J. A polarizing situation: Taking an in-plane perspective for next-generation near-field studies. *Frontiers of Physics* **11**, doi:10.1007/s11467-015-0526-5 (2016).
36  Haigh, S. J. *et al.* Cross-sectional imaging of individual layers and buried interfaces of graphene-based heterostructures and superlattices. *Nature Materials* **11**, 764, doi:10.1038/nmat3386 (2012).
37  Shi, W. *et al.* Raman and photoluminescence spectra of two-dimensional nanocrystallites of monolayer $WS_2$ and $WSe_2$. *2D Materials* **3**, doi:10.1088/2053-1583/3/2/025016 (2016).
38  Eggleston, M. S., Messer, K., Zhang, L., Yablonovitch, E. & Wu, M. C. Optical antenna enhanced spontaneous emission. *Proceedings of the National Academy of Sciences* **112**, 1704-1709 (2015).
39  Zhang, X. X., You, Y., Zhao, S. Y. & Heinz, T. F. Experimental Evidence for Dark Excitons in Monolayer $WSe_2$. *Physical Review Letters* **115**, 257403, doi:10.1103/PhysRevLett.115.257403 (2015).
40  Park, K. D., Jiang, T., Clark, G., Xu, X. D. & Raschke, M. B. Radiative control of dark excitons at room temperature by nano-optical antenna-tip Purcell effect. *Nature Nanotechnology* **13**, 59-+, doi:10.1038/s41565-017-0003-0 (2018).
41  Chow, P. K. *et al.* Defect-Induced Photoluminescence in Monolayer Semiconducting Transition Metal Dichalcogenides. *ACS Nano* **9**, 8 (2015).
42  Kern, J. *et al.* Nanoscale Positioning of Single-Photon Emitters in Atomically Thin $WSe_2$. *Advanced Materials* **28**, 7101-7105, doi:10.1002/adma.201600560 (2016).
43  Luo, Y. *et al.* Deterministic coupling of site-controlled quantum emitters in monolayer $WSe_2$ to plasmonic nanocavities. *Nat Nanotechnol* **13**, 1137-1142, doi:10.1038/s41565-018-0275-z (2018).
44  Chiang, C. L., Xu, C., Han, Z. M. & Ho, W. Real-space imaging of molecular structure and chemical bonding by single-molecule inelastic tunneling probe. *Science* **344**, 885-888, doi:10.1126/science.1253405 (2014).
45  van Duin, A. C. T.; Dasgupta, S.; Lorant, F.; Goddard, W. A. *J. Phys. Chem. A* 2001, *105*, 9396–9409, DOI: 10.1021/jp004368u
46  Ostadhossein, A.; Rahnamoun, A.; Wang, Y.; Zhao, P.; Zhang, S.; Crespi, V. H.; van Duin, A. C. T. *J. Phys. Chem. Lett.* 2017, *8*, 631–640, DOI: 10.1021/acs.jpclett.6b02902
47  Froyen, S.; Harrison, W. A. *Phys. Rev. B: Condens. Matter Mater. Phys.* 1979, *20*, 2420–2422, DOI: 10.1103/PhysRevB.20.2420